\documentclass[amsmath,twocolumn,pra]{revtex4}
\usepackage{dcolumn}    
\usepackage{bm}         
\usepackage{ifpdf}
\usepackage{amssymb,lineno,amsfonts}
\usepackage{graphicx}   
\usepackage{bbm}
\usepackage{mathrsfs}
\usepackage{upgreek}
\usepackage{epstopdf}
\usepackage{setspace}
\usepackage{hyperref}
\usepackage[matrix,frame,arrow]{xypic}
\usepackage{rotating}
\usepackage{float}
\usepackage{natbib}
\usepackage{multirow}

\newcommand{\ket}[1]{\vert{#1}\rangle}

\newcommand{\outpr}[2]{\vert{#1}\rangle\langle{#2}\vert}

\begin{document}

\title{Pauli Decomposition over Commuting Subsets: \\ Applications in Gate Synthesis, State Preparation, and Quantum Simulations}
\author{Swathi S. Hegde$^1$, K. R. Koteswara Rao$^2$ and T. S. Mahesh$^1$}
\email{mahesh.ts@iiserpune.ac.in}
\affiliation{$^1$Department of Physics and NMR Research Center, 
Indian Institute of Science Education and Research, Pune 411008, India
\\
$^2$Fakult\"{a}t Physik, Technische Universit\"{a}t Dortmund, D-44221 Dortmund, Germany}

\begin{abstract} 
A key task in quantum computation is the application of
a sequence of gates implementing a specific unitary operation.  However, the decomposition of an arbitrary 
unitary operation into simpler quantum gates is a nontrivial problem.  Here we propose a general and robust protocol to decompose any target unitary into a sequence of Pauli rotations.  The procedure involves identifying a commuting subset of Pauli operators having a high trace overlap with the target unitary, followed by a numerical optimization of their corresponding rotation angles.  The protocol is demonstrated by decomposing several standard quantum operations.  The applications of the protocol for quantum state preparation and quantum simulations are also described.  Finally, we describe an NMR experiment implementing a three-body quantum simulation, wherein the above decomposition technique is used for the efficient realization of propagators.
\end{abstract}
\maketitle

\section{Introduction}
Quantum devices have the capability to perform several tasks with efficiencies beyond the reach of their classical counterparts \cite{chuang,preskill}.  An important criterion for the physical realization of such devices is to achieve precise control over the quantum dynamics \cite{divincenzo}. 
The circuit model of quantum computation is based on the  realization of a desired unitary in terms of simpler quantum gates.  However, arbitrarily precise decomposition of a general unitary $U_T$ in the form 
\begin{equation}
U_T = U_m \cdots U_2 U_1,
\label{U}
\end{equation}
is a nontrivial task.
Here each of the $U_j$'s is either of lower complexity or acts on smaller subsystems. Such a decomposition is said to be efficient if (i) $m$ scales polynomially with the system size $n$ and (ii) spatial and temporal overhead of each $U_j$ scales polynomially with $n$.

The decomposition of a unitary operator corresponding to a Hamiltonian ${\cal H} = {\cal H_A}+{\cal H_B}$, where $[{\cal H_A},{\cal H_B}] = 0$, is
trivial, i.e., $U_T = e^{-i{\cal H} t} = e^{-i{\cal H_A} t}e^{-i{\cal H_B} t}$.   When  $[{\cal H_A},{\cal H_B}] \neq 0$, 
one can discretize the time, $\delta = t/m$,
and use the Trotter's formula \cite{trotter}
\begin{equation}
U_T = \left[e^{-i{\cal H_A} \delta}e^{-i{\cal H_B}\delta}\right]^m + {\cal O}(\delta^2)
\label{U_trot}
\end{equation}
 or
its symmetrized form \cite{trotter1}
\begin{equation}
U_T = \left[e^{-i{\cal H_A} \delta/2}e^{-i{\cal H_B} \delta}e^{-i{\cal H_A} \delta/2}\right]^m  + {\cal O}(\delta^3)
\label{U_trots}.
\end{equation}
For a time-dependent Hamiltonian ${\cal H}(t)$, one needs to use Dyson's time ordering operator or the Magnus expansion, and then decompose the time discretized components \cite{sakurai2011modern}.
However, for a given unitary $U_T$, such a decomposition may not be obvious or even after the decomposition, the individual pieces themselves may involve matrix exponentials of non-commuting operators thus failing to reduce the complexity. 

Several advanced decomposition routines have been suggested for arbitrary unitary decomposition. Barenco {\em et al.} have shown that XOR gates along with local gates are universal, and in terms of these elementary gates they have explicitly decomposed several standard quantum operations \cite{barenco}. Tucci presented an algorithm to decompose an arbitrary unitary into single and two-qubit gates using a mathematical technique called CS decomposition \cite{tucci1999rudimentary}. Khaneja {\em et al.} used Cartan decomposition of the semi-simple lie group SU$(2^n)$ for the unitary decomposition \cite{khaneja2001cartan}. A method to realize any multiqubit gate  using fully controlled single-qubit gates using Grey code was given by Vartiainen {\em et al.} \cite{vartiainen2004efficient}.
M\"{o}tt\"{o}nen {\em et al.} have presented a cosine-sine matrix decomposition method synthesizing the gate sequence
to implement a general $n$-qubit quantum gate \cite{mottonen2004quantum}.
 Recently, Ajoy {\em et al.} also developed an ingenious algorithm to 
decompose an arbitrary unitary operator 
using algebraic methods \cite{ashok}. 
More recently, this method was utilized in the experimental implementation of mirror inversion and other quantum state transfer protocols \cite{kota}.
Manu et al have shown several unitary decompositions by case-by-case numerical optimizations \cite{manu}.

In this article, we propose a general algorithm to decompose an arbitrary unitary upto a desired precision.
It is distinct from the above approaches in several ways. 
Firstly, our method considers generalized rotations of commuting Pauli operators which are more amenable for practical implementations via optimal control techniques.  Secondly, being a numerical procedure, it considers various experimentally relevant parameters such as robustness with respect to fluctuations in the control parameters, minimum rotation angles etc. Besides, the procedure can be extended for quantum circuits, quantum simulations, quantum state preparations, and probably in some cases even for nonunitary synthesis.

The paper is organized as follows: A detailed explanation of the algorithm for arbitrary unitary decomposition is presented in section II .  Section III deals with the applications of the algorithm with explicit demonstrations involving the decomposition of standard gate-synthesis, quantum circuit designs, certain quantum state preparations, and for quantum simulations. Finally we conclude in section IV.

\begin{figure}
	\includegraphics[trim=4.5cm 0cm 4.5cm 0cm, clip=true, width=9cm]{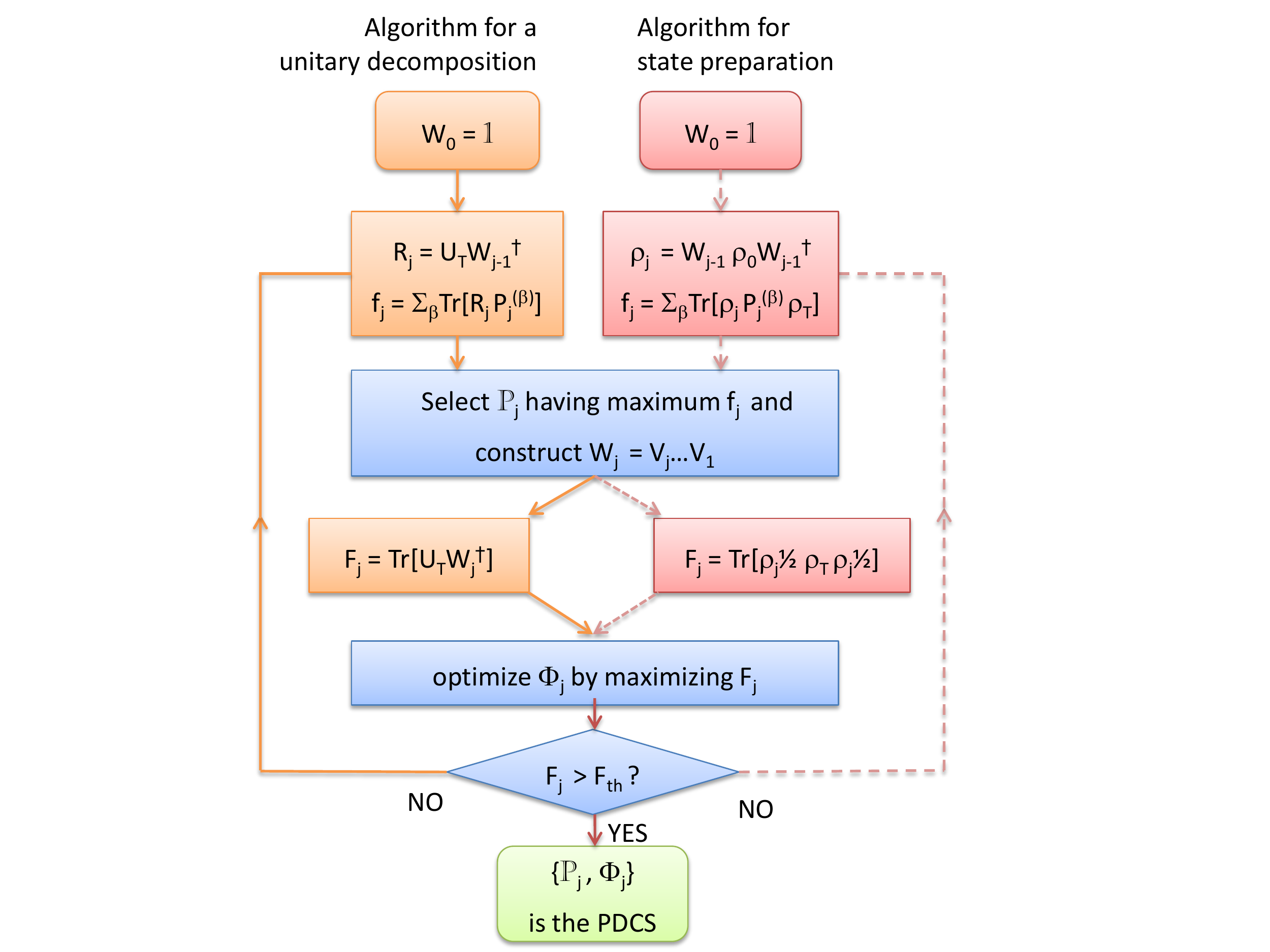}
	\caption{(Color online) The flowchart describing PDCS algorithm for unitary decomposition (left) and state preparation (right, dashed).
	}
	\label{flowchart}
\end{figure}

\begin{figure}[b]
	\includegraphics[trim=0.2cm 17.5cm 6.5cm 0.5cm, clip=true, width=15cm]{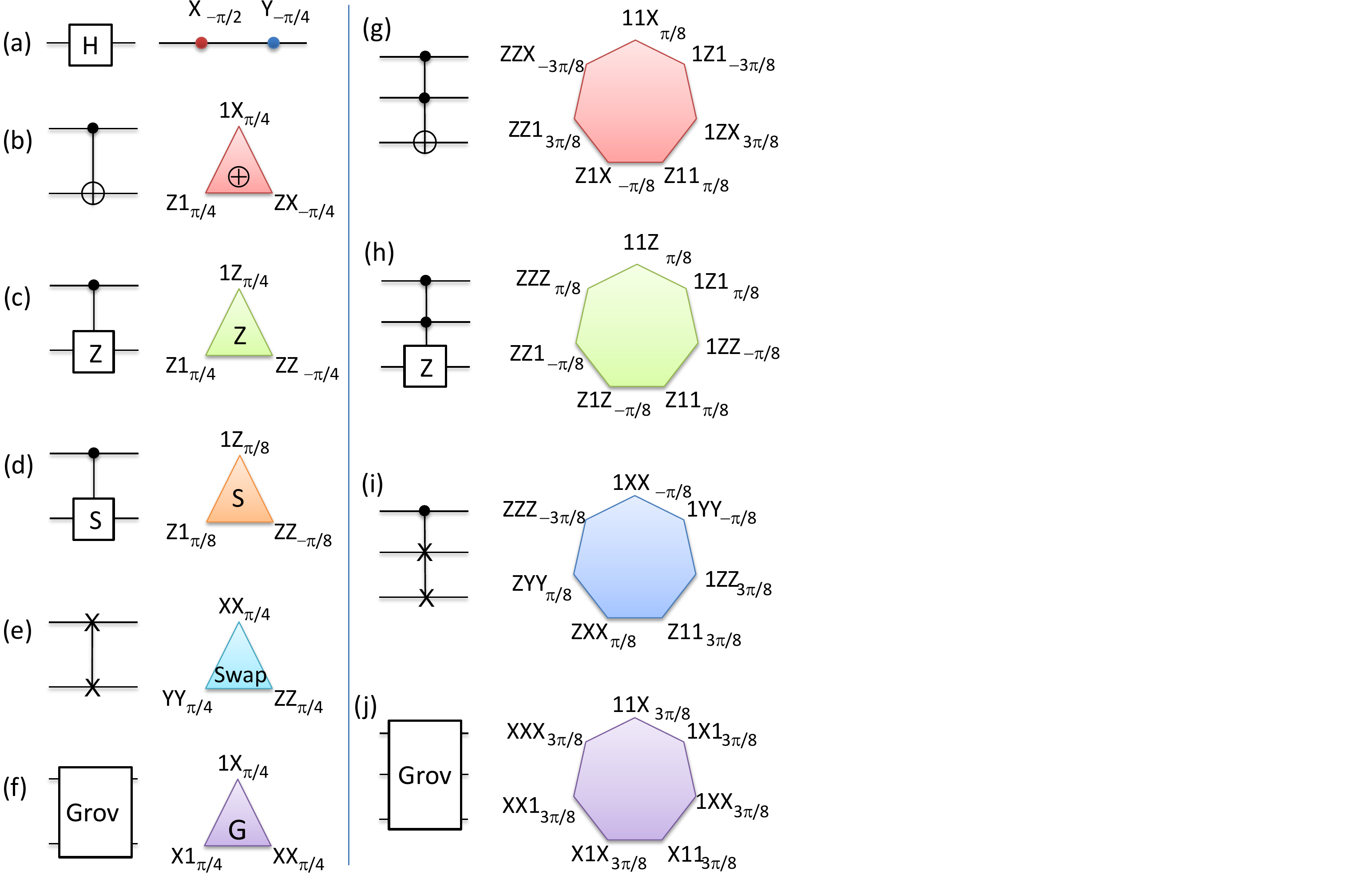}
	\caption{(Color online) PDCS of some standard quantum gates: (a) single-qubit Hadamard, (b) c-NOT, (c) c-Z,
		(d) c-S, (e) SWAP, (f) 2-qubit Grover iterate, (g) Toffoli, (h) c$^2$-Z, (i) Fredkin, and (j) 3-qubit Grover iterate.  The individual rotors are represented by dots, triangles, and heptagons depending on number of Pauli operators (indicated at the vertices) in each rotor. The corresponding rotation angles are indicated by subscripts.
		\label{gates1}}
\end{figure}

\section{Algorithm}
In the following, we describe an algorithm for Pauli decomposition over commuting subsets (PDCS) for arbitrary unitary operators. Although for the sake of simplicity we utilize 2-level quantum systems,  the protocol applies equally well to any $d$-level quantum systems.

Let $U_T$ be the desired unitary operator of dimension $N=2^n$ to be applied on an $n$-qubit system.  We seek an $m$-rotor decomposition
$W_m = V_m V_{m-1} \cdots V_1 \approx U_T$,
where the decomposed unitaries $V_j$ with $j\in[1,m]$ have the form
\begin{equation}
	V_j = e^{-i \sum_\beta P^{(\beta)}_j \phi^{(\beta)}_j }.
	\label{Uj}
\end{equation}
Here $\{P^{(\beta)}_j\} \equiv \mathbb{P}_j$ is a maximally commuting subset of $n-$ qubit Pauli operators,  $\{\phi^{(\beta)}_j\}\equiv \Phi_j$ is the set of corresponding rotation angles, and the index $\beta$ runs over the elements of $\mathbb{P}_j$.
In general, for an $n$-qubit case, a maximal commuting subset $\mathbb{P}_j$ can have at most $N-1$ elements. 
The fidelity $F_m$ of the decomposition, defined by
\begin{equation}
F_m = \langle U_T \vert W_m \rangle = \vert \mathrm{Tr}[U_T^\dagger W_m]/N \vert ,
\label{fd}
\end{equation}
should be larger than a desired threshold $F_\mathrm{th}$.

\begin{figure*}
	\includegraphics[trim=0.2cm 24.8cm 5cm 1.5cm, clip=true, width=15cm]{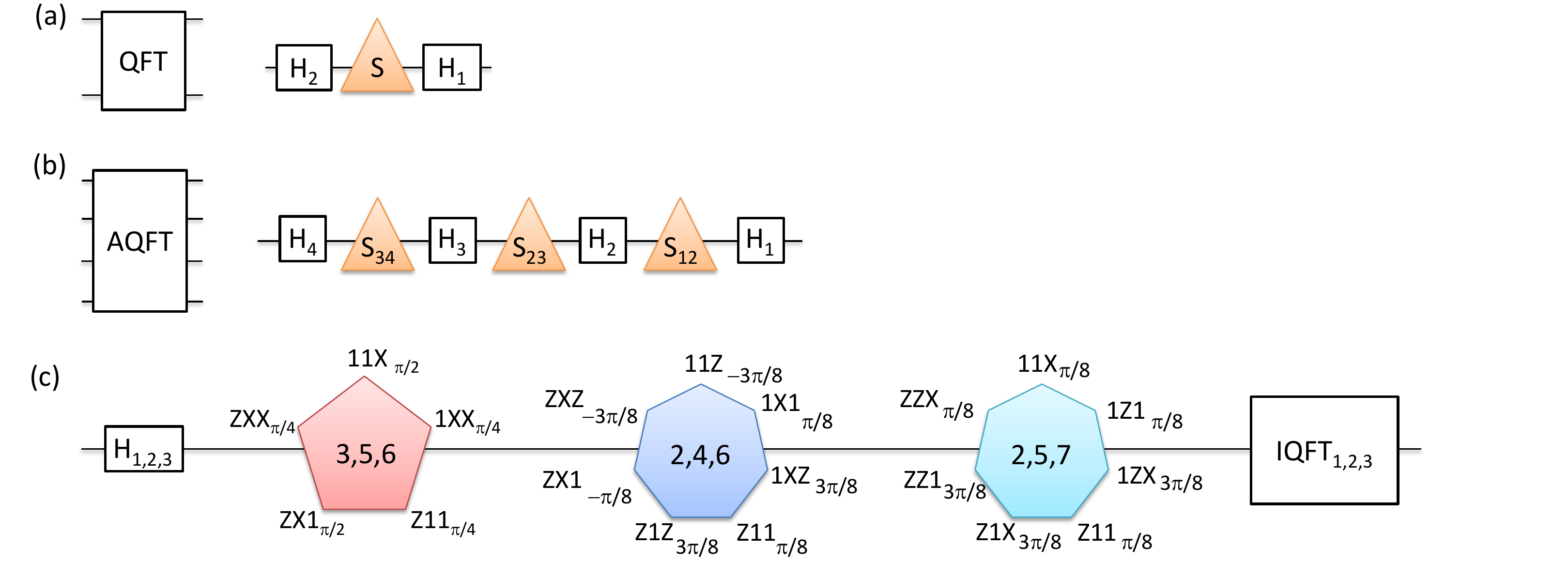}
	\caption{(Color online) 
		PDCS of (a) 2-qubit Quantum Fourier Transform (QFT), 
		(b) 4-qubit approximate QFT (AQFT), and
		(c) 7-qubit Shor's circuit for factorizing 15.
		In (b), each S gate acts on a pair of qubits as indicated by the subscripts.
		\label{gates2}}
\end{figure*}

The flowchart for the PDCS algorithm is shown in Fig. \ref{flowchart}.
We now describe an algorithm to build $W_m$ in $m$ steps.  
To begin with, we start with $W_0 = \mathbbm{1}$.
The $j$th step of the algorithm consists of the following processes:
\begin{enumerate}
\item
Calculate the residual propagator 
$R_{j} = U_T W_{j-1}^\dagger$. 
\item Selection of the commuting subset $\mathbb{P}_j$ having the maximum overlap 
$f_j = \sum_\beta \mathrm{Tr} [R_{j} P^{(\beta)}_j]$
with the residual unitary $R_j$.
\item Setting up the decomposition $W_j$
and numerically optimizing the rotation angles $\{\Phi_1,\cdots,\Phi_j\}$ by maximizing the fidelity $F_j = \langle W_j \vert U_T\rangle$, where $W_j = V_j \cdots V_1$.
\end{enumerate}
These steps can be iterated upto $m$-steps until the fidelity $F_m > F_\mathrm{th}$ of a desired value is reached.

In general, the solutions to the decomposition may not be unique.  However, it is desirable to attain a decomposition that is most suitable for experimental implementations.  In this regard, we look for solutions with minimum rotation angles $\{\Phi_j\}$, which can be obtained by using a suitable penalty function in step 3 of the above algorithm.

\section{Applications}
\subsection{PDCS of quantum gates and circuits}
In this section we illustrate PDCS of several standard quantum gates. As described in Eq. \ref{Uj}, the $j$th decomposition is expressed in terms of the commuting Pauli operators $\mathbb{P}_j$ and the corresponding rotation angles $\Phi_j$.  Further, a specific operator $P_j^{(\beta)} \in \mathbb{P}_j$ can be expressed as a tensor product of single-qubit Pauli operators $X$, $Y$, $Z$, and the identity matrix $\mathbbm{1}$.

Exact PDCS of several standard quantum gates are shown in Fig. \ref{gates1}.  For a single-qubit Hadamard operation (Fig. \ref{gates1}(a)), we obtain a decomposition with two noncommuting rotations, as is well known \cite{chuang}.  Here $\mathbb{P}_1=X,~ \mathbb{P}_2 = Y$ and the corresponding rotation angles $\Phi_1 = -\pi/2, ~ \Phi_2 = -\pi/4$,  are indicated by the subscripts.

In the two-qubit case, the maximal commuting subset can have only three Pauli operators and there are only 15 such subsets. Figs. \ref{gates1}(b-e) describe decompositions of several two-qubit gates namely c-NOT, c-Z, c-S, and SWAP gate. 
Here	$Z = \left[
		\begin{array}{cc}
				1 & 0 \\
				0 & -1
		\end{array}
		\right]$ 
		and
		$S = \left[
		\begin{array}{cc}
		1 & 0 \\
		0 & i
		\end{array}
		\right]$.
It is interesting to note that each of these gates needs a single subset of commuting Pauli operators.  Such a rotation can be obtained by a single matrix exponential and can be thought of as a single generalized rotation in the Pauli space. We refer to such a generalized rotation as a rotor, and since it consists of three operators, we represent it by a triangle. In practice, the individual components of a single rotor can be implemented either simultaneously, or in any order. We find that even a 2-qubit Grover iterate, i.e., $G = 2\outpr{\psi}{\psi}-\mathbbm{1}$, 
where $\ket{\psi} = (\ket{00}+\ket{01}+\ket{10}+\ket{11})/2$, can be realized as a single rotor (Fig. \ref{gates1}(f)).

For three-qubits, the maximal commuting subset can have only seven commuting Pauli operators and there are 135 distinct subsets.  Figs. \ref{gates1}(g-j) describe Toffoli, c$^2$-Z, Fredkin, and 3-qubit Grover iteration respectively.  Again we find that a single heptagon rotor suffices for realizing each of the standard gates.  Similarly, in the case of four-qubits, a maximal commuting subset can have 15 operators and one can verify that a basic gate such as c$^3$-NOT, c$^3$-Z, etc. can be realized by a single rotor.  

It is always possible to decompose a multi-qubit quantum circuit in terms of single- and two-qubit gates \cite{barenco}.  
As  examples, we shall consider PDCS of a few quantum circuits (Fig. \ref{gates2}). The two-qubit Quantum Fourier Transform (QFT) circuit can be exactly decomposed into three rotors as shown in Fig. \ref{gates2}(a). As another example, PDCS of the 4-qubit approximate QFT (AQFT) circuit \cite{aqft}  results in only single-qubit and two-qubit rotors as shown in Fig. \ref{gates2}(b).  Similarly, PDCS of the 7-qubit Shor's circuit for factoring the number 15  involves at most three-qubit rotors as shown in Fig. \ref{gates2}(c) \cite{shor}.  In this sense, PDCS of multiqubit quantum gates and quantum circuits is scalable with increasing system size.

\begin{figure}[t]
	\includegraphics[trim=2cm 25.5cm 18cm 1.5cm, clip=true, width=8.5cm]{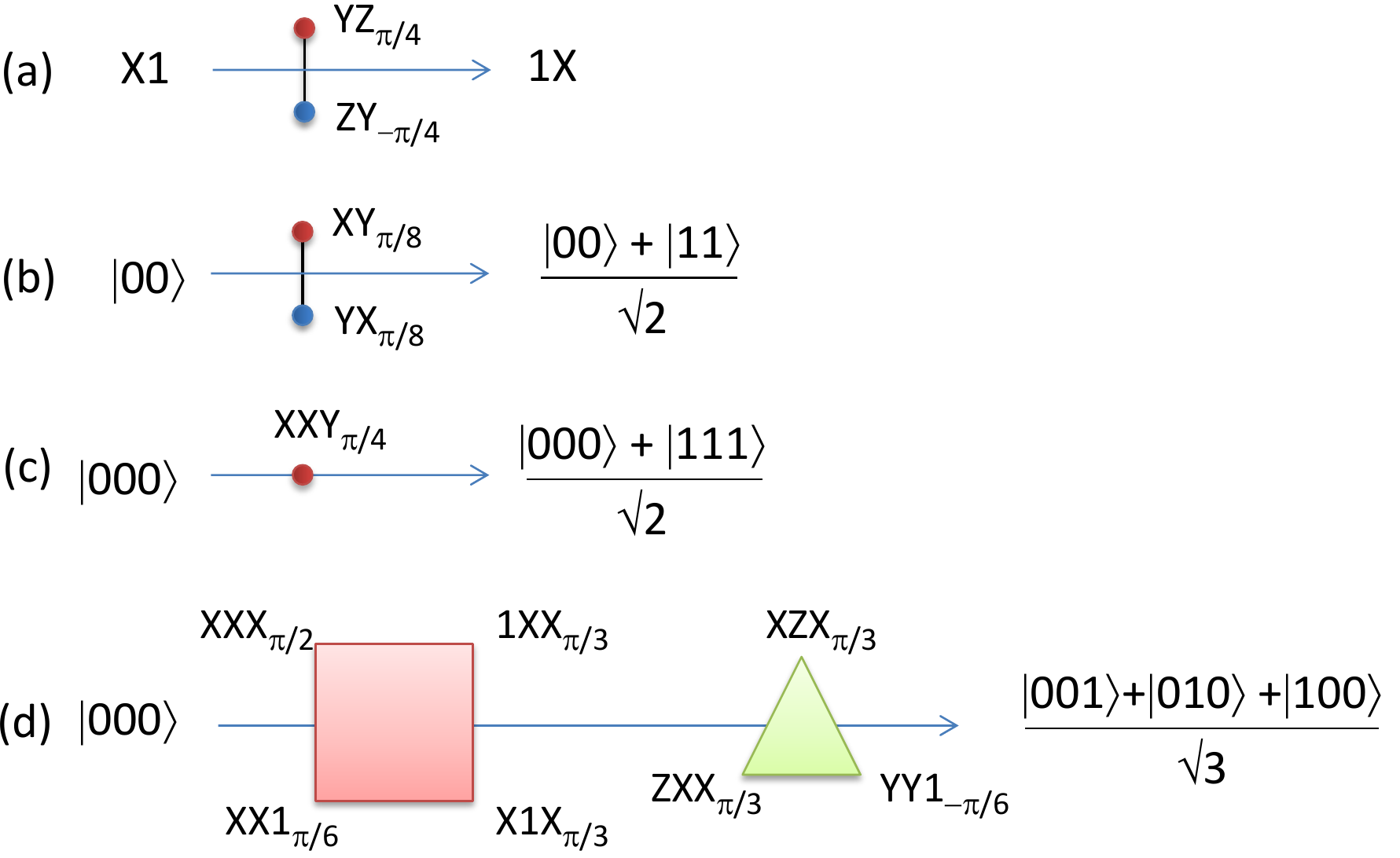}
	\caption{(Color online) PDCS of some state to state transfers: (a) polarization transfer (INEPT) and (b-d) preparation of Bell, GHZ, and W states respectively starting from pure states.}
	\label{pdstate}
\end{figure}

\subsection{Quantum state preparation}
Here the goal is to prepare a target state
$\rho_T$ starting from a given initial state
$\rho_0$.  In general the unitary operator, connecting the initial and target states, itself is not unique.  The procedure is similar to that described in the previous section, and is summarized in the flowchart shown in Fig. \ref{flowchart}. Here 
the selection of commuting Pauli operators $\mathbb{P}_j$ is based on the overlap 
$f_j = \sum_{\beta} \mathrm{Tr}[\rho_j P_j^{(\beta)} \rho_T]$,
where $\rho_j = W_j \rho_0 W_j ^\dagger$ is the intermediate state after $j$th decomposition.  As explained before, we select
the commuting subset $\mathbb{P}_j$ having the maximum overlap $f_j$ and optimize the phases $\Phi_j$ by maximizing the Uhlmann fidelity 
\begin{eqnarray}
F_j = \langle \rho_j \vert \rho_T \rangle =  \mathrm{Tr}[\rho_j^{1/2}\rho_T\rho_j^{1/2}].
\end{eqnarray}
Again, $m$ iterations are carried out until  $F_m \ge F_\mathrm{th}$ is realized.

Fig. \ref{pdstate} displays PDCS of some standard state to state transfers.  The polarization transfer in a pair of qubits (popularly known as INEPT \cite{cavanagh}) requires a single rotor having a pair of bilinear operators (Fig. \ref{pdstate}(a)).
The preparation of a Bell and GHZ states respectively from $\ket{00}$ and $\ket{000}$ states also require a single rotor (Fig. \ref{pdstate}(b-c)).  However, the preparation of a three-qubit W-state is somewhat more elaborated, and requires two rotors (Fig. \ref{pdstate}(d)).  Although these decompositions are not unique, it is possible to optimize them based on the experimental conditions.
Here one can notice that although a maximal commuting subset can have up to $N-1$ elements, it is often possible to decompose a multi-qubit operation over a smaller commuting subset.

\begin{figure}[b]
	\includegraphics[trim=4cm 8.5cm 2cm 4.5cm, clip=true, width=8cm]{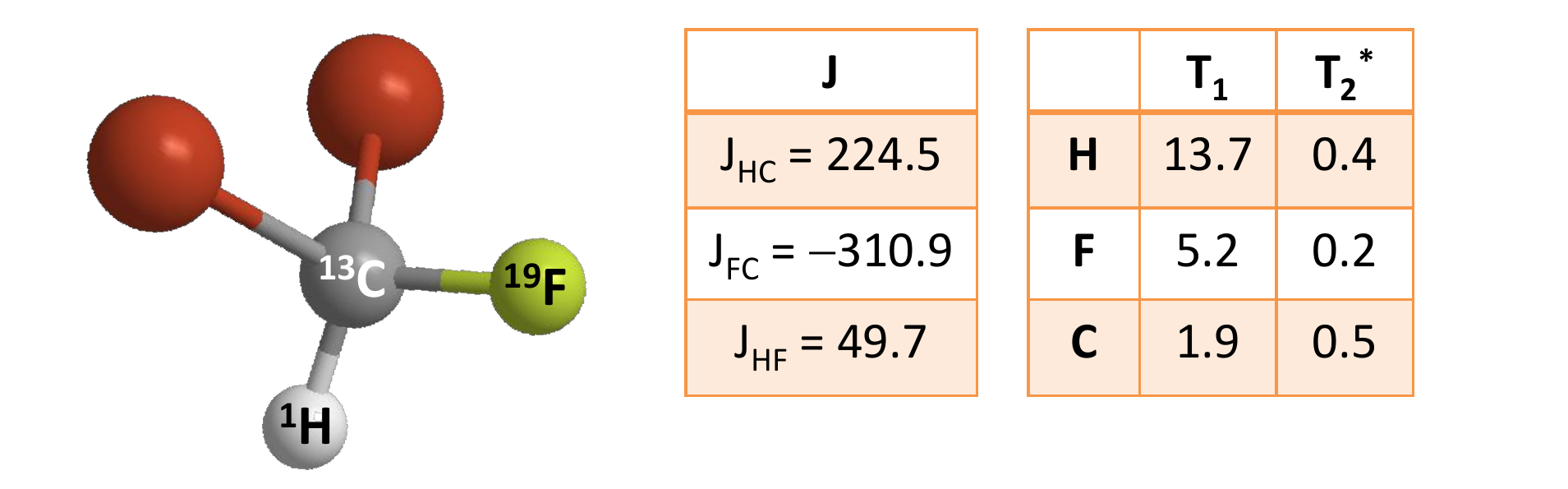}
	\caption{(Color online) Dibromofluoromethane consisting of three nuclear spin qubits $^1$H, $^{13}$C and $^{19}$F. The tables display the values of indirect spin-spin coupling constants ($J$) in Hz, and the relaxation time constants ($T_1$, and $T_2^*$) in seconds.}
	\label{hfc}
\end{figure}

\subsection{Quantum simulation}
Utilizing controllable quantum systems to mimic the dynamics of other quantum systems is the essence of quantum simulation \cite{feynman}. Various quantum devices have already demonstrated quantum simulations of a number of quantum mechanical phenomena (for example, \cite{suter2005,nori,atoms1,freeze}). 
An important application of the decomposition technique described above is in the experimental realization of quantum simulations.
To illustrate this fact,  we experimentally carryout  quantum simulation of a three-body interaction Hamiltonian using a three-qubit system. While such a
Hamiltonian is physically unnatural, simulating such interactions has interesting applications such as in quantum state transfer \cite{suter_3body_statetransfer}.  Specifically,
we simulate the dynamics under the Hamiltonian
\begin{eqnarray}
{\cal H}_\mathrm{S} = X\mathbbm{1}\mathbbm{1}+\mathbbm{1}X\mathbbm{1}+\mathbbm{1}\mathbbm{1}X + J_{123} ZZZ,
\label{H_3body}
\end{eqnarray}
where $ J_{123}$ is the three-body interaction strength.  
A slightly different three-body Hamiltonian simulated earlier by Cory and coworkers \cite{3body} consisted of only the last term.  The presence of other terms which are noncommuting with the 3-body term necessitates an efficient decomposition of the overall unitary.

We use three spin-1/2 nuclei of dibromofluoromethane (Fig. \ref{hfc}) dissolved in acetone-D6 as our three-qubit system.  All the 
experiments are carried out on a 500 MHz Bruker NMR spectrometer at an ambient temperature of
300 K.  In the triply rotating frame at resonant offsets, the internal Hamiltonian of the system is given by
\begin{eqnarray}
{\cal H}_\mathrm{int} = 
\left[ J_\mathrm{HF}ZZ\mathbbm{1} +
J_\mathrm{HC}Z\mathbbm{1}Z +
J_\mathrm{FC}\mathbbm{1}ZZ\right]\pi/2,
\end{eqnarray}
and the values of the indirect coupling constants $\{J_\mathrm{HF},J_\mathrm{HC},J_\mathrm{FC}\}$ are as in Fig. \ref{hfc}.  This internal Hamiltonian along with the external control Hamiltonians provided by the RF pulses are used in the following to mimic the three-body Hamiltonian in Eqn. \ref{H_3body}.

\begin{figure}[b]
	\includegraphics[trim=6.5cm 8.2cm 4.5cm 5cm, clip=true, width=6.5cm]{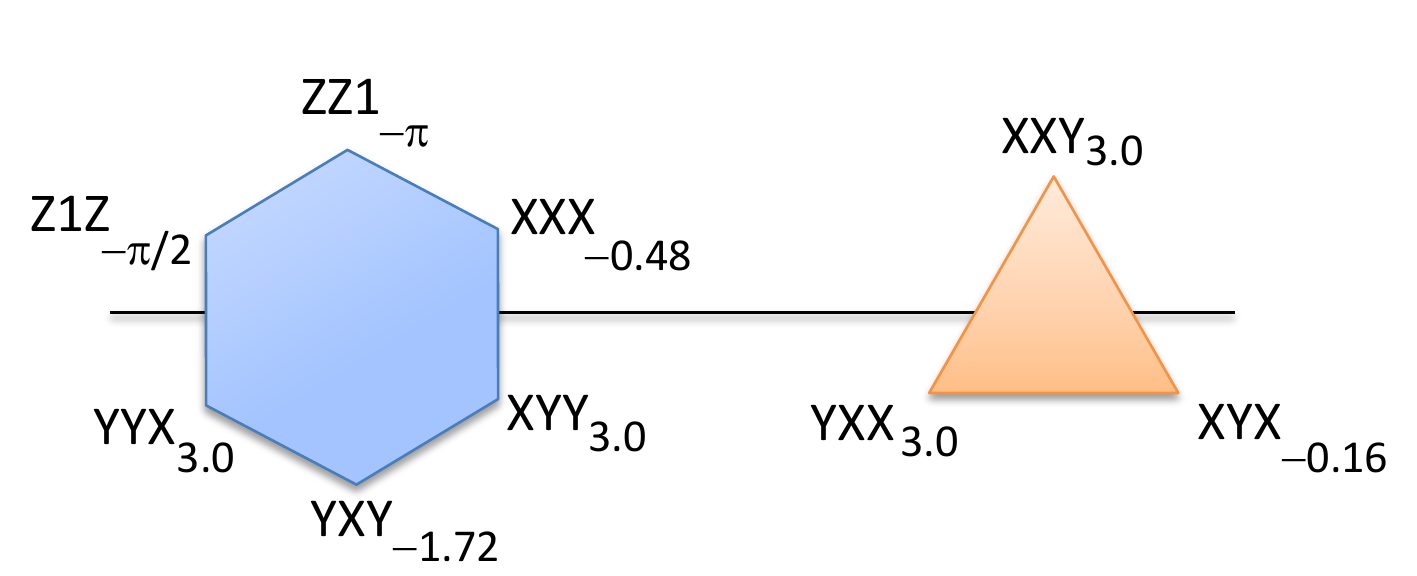}
	\caption{(Color online) PDCS of 
		$U_s(\tau) = \exp(-i{\cal H}_s\tau)$ for $J_{123}=5$ Hz and $\tau = 0.8$ s.}
	\label{U_3body}
\end{figure}

The traceless part of the thermal equilibrium state of the 3-qubit NMR spin system is given by
$\rho_{eq} = (Z\mathbbm{11}+\mathbbm{1}Z\mathbbm{1}+\mathbbm{11}Z)/2$ \cite{cavanagh}.  The initial state $\rho(0) = (X\mathbbm{11}+\mathbbm{1}X\mathbbm{1}+\mathbbm{11}X)/2$ is prepared by applying a $90^\circ$ RF-pulse  about $Y$.  
The goal is to subject the three-spin system to an effective three-body Hamiltonian $H_S$ and monitor the evolution of its state
$\rho(t) = U_S(t) \rho(0) U_S(t)^\dagger$, where
$U_S(t) = e^{-i{\cal H}_St}$.
We choose to experimentally observe the  transverse magnetization 
\begin{eqnarray}
M_x(k \tau) = \mathrm{Tr}[\rho(k \tau)(X\mathbbm{11}+\mathbbm{1}X\mathbbm{1}+\mathbbm{11}X)/2]
\end{eqnarray}
for $ J_{123}=5$ Hz at discrete time intervals $k \tau$, where $k = \{0,\cdots,20\}$ and $\tau=0.8$ s.

 \begin{figure}[t]
 	\includegraphics[trim=0.2cm 8.9cm 0cm 0cm, clip=true, width=9cm]{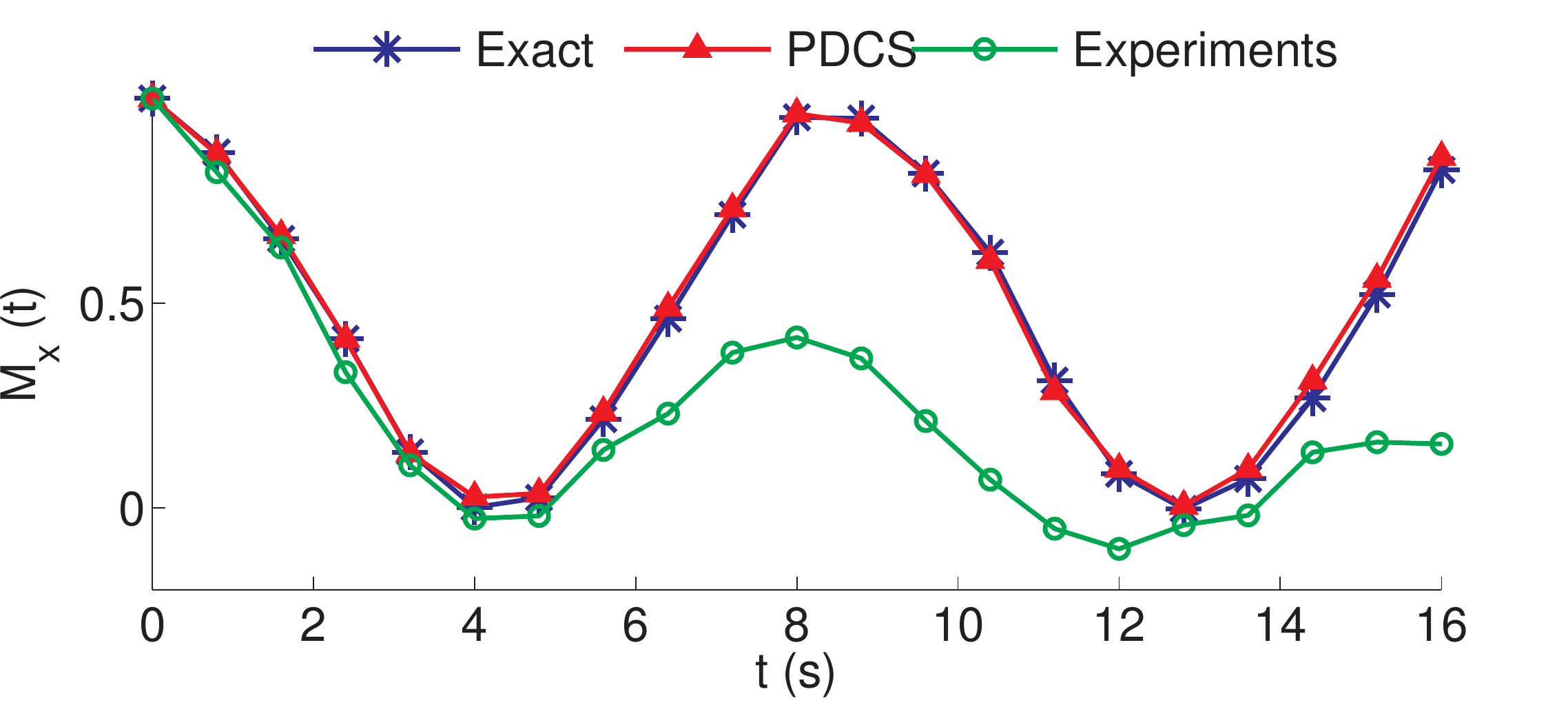}
 	\caption{(Color online) Magnetization vs time. 
 	}
 	\label{simu}
 \end{figure}

The PDCS of $U_S$ shown in Fig. \ref{U_3body} consists of two rotors: a hexagon followed by a triangle.  We utilized bang-bang (BB) control technique for generating each of the two rotors \cite{gaurav}.  The duration of each BB-sequence was about 7 ms and fidelities were above 0.98 averaged over a 10\% inhomogeneous distribution of RF amplitudes.  The results of the experiments (hollow circles) and their comparison with numerical simulation of the PDCS (triangles) and exact numerical values (stars) are shown in Fig. \ref{simu}.  The first experimental data point was obtained after a simple 90 degree RF pulse and was normalized to 1.
The $k^{\mathrm{th}}$ point is obtained by $k$ iterations of the BB-sequence for $U_S(\tau)$. While the experimental curve displays the same period and phase as that of the simulated curve, the steady decay in amplitude is mainly due to decoherence and other experimental imperfections such as RF inhomogeneity and  nonlinearities of the RF channel.

\begin{figure}[b]
	\includegraphics[trim=3cm 5cm 3cm 5cm, clip=true, width=8cm]{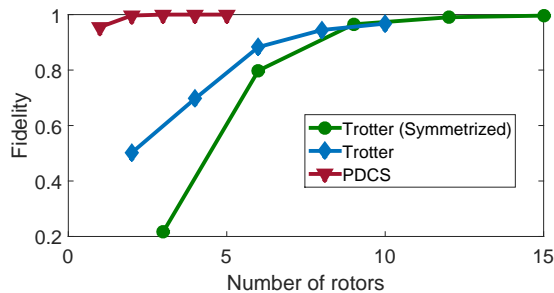}
	\caption{(Color online) Fidelity ($F_m$) vs number ($m$) of rotors in the decompositions. 
	}
	\label{comptrot}
\end{figure}

To compare the efficiency of PDCS with that of Trotter decomposition (in Eq. \ref{U_trot} and \ref{U_trots}), we calculate the fidelities ($F$) of the decomposed propagator with the exact propagator
$U_S(\tau)$ as a function of number $m$ of rotors (see Fig. \ref{comptrot}).  It can be observed that, with increasing number of rotors, PDCS fidelity converges faster than the Trotter.

\section{Conclusions}
In this work we proposed a generalized numerical algorithm based on Pauli decomposition over commuting subsets (PDCS). The aim of the algorithm is to decompose an arbitrary target unitary into simpler unitaries, referred to as rotors.  Each rotor consists of only commuting subset of Pauli operators. These rotors are optimized to be robust against experimental errors by minimizing the rotation angles and by considering other control errors.  Thus apart from providing an intuitive and topological representation of an arbitrary quantum circuit, the method is also useful for its efficient physical  realization.

We demonstrated the robustness and efficiency of the decomposition using numerous examples of quantum gates and circuits. It is interesting to note that several standard quantum gates correspond to single rotors.     
We also discussed the applications of  PDCS in quantum state-to-state transfers and illustrated it using several examples. 

Another important application of PDCS is in  quantum simulations.  As an 
example, we described the quantum simulation of a three-body interaction.  We used PDCS algorithm to decompose the unitary corresponding to such a Hamiltonian and found it to be more efficient than Trotter decomposition
in terms of the number of rotors. Further, we have demonstrated the quantum simulation by experimentally monitoring the evolution of magnetization using a three qubit NMR system.  The experimental results matched with the numerical simulations upto a decay factor arising  predominantly due to decoherence. 

 We believe such unitary decomposition strategies combined with sophisticated optimal control techniques will greatly assist in efficient quantum control. \\

\section*{Acknowledgments}
We thank Prof. Anil Kumar, IISc, Bangalore for providing us the three-qubit NMR register. 
We also acknowledge useful discussions with Sudheer Kumar.
This work was partly supported by DST Project No. DST/SJF/PSA-03/2012-13.

\bibliographystyle{apsrev4-1} 
\bibliography{pd1}

\end{document}